\documentclass[prl,twocolumn,floatfix,superscriptaddress]{revtex4}

\usepackage{amsmath,amssymb,amsfonts,xspace}
\usepackage{epsfig,supertabular}
\usepackage{dsfont}

\def\det{{\rm det}\, }

\def\({\left(}
\def\){\right)}
\def\[{\left[}
\def\]{\right]}

\newcommand{\de}{\mathrm{d}}

\newcommand{\I}{\mathrm{i}}

\newcommand{\cF}{\mathcal{F}}

\newcommand{\cM}{\mathcal{M}}
\newcommand{\cN}{\mathcal{N}}
\newcommand{\cD}{\mathcal{D}}

\newcommand{\cP}{\mathcal{P}}
\newcommand{\cR}{\mathcal{R}}

\DeclareSymbolFont{AMSa}{U}{msa}{m}{n}
\DeclareSymbolFont{AMSb}{U}{msb}{m}{n}
\DeclareMathSymbol{\fieldR}{\mathalpha}{AMSb}{"52}

\newcommand{\cO}{\mathcal{O}}

\newcommand{\IR}{\mathbb{R}}

\newcommand{\IZ}{\mathbb{Z}}

\newcommand{\Tr}{{\rm Tr}}

\def\bea{\begin{eqnarray}}
\def\eea{\end{eqnarray}}
\def\be{\begin{equation}}
\def\ee{\end{equation}}
\def\ba{\begin{align}}
\def\ea{\end{align}}
\def\bse{\begin{subequations}}
\def\ese{\end{subequations}}

\begin{document}

\title{\

\vspace{-13mm}
{\begin{normalsize}
\mbox{} \hfill {\rm CPHT-RR-041-0816, CERN-TH-2016-175} \\
\vskip 10pt
\end{normalsize}}

{\bf\Large
Protected couplings and BPS dyons  in half-maximal \\[1mm]
supersymmetric string vacua 
} }

\author{Guillaume Bossard}
\email{guillaume.bossard@polytechnique.edu}
\affiliation{Centre 
de Physique Th\'eorique, Ecole Polytechnique, Universit\'e Paris-Saclay, 91128 Palaiseau Cedex, France}

\author{Charles Cosnier-Horeau}
\email{charles.cosnier-horeau@polytechnique.edu}
\affiliation{Centre 
de Physique Th\'eorique, Ecole Polytechnique, Universit\'e Paris-Saclay, 91128 Palaiseau Cedex, France}
\affiliation{Sorbonne Universit\'eŽs, Laboratoire de Physique Th\'eorique et Hautes
Energies, CNRS UMR 7589,
Universit\'e Pierre et Marie Curie,
4 place Jussieu, 75252 Paris Cedex 05, France}

\author{Boris Pioline}
\email{boris.pioline@cern.ch}
\affiliation{CERN, Theoretical Physics Department, 1211 Geneva 23, Switzerland}
\affiliation{Sorbonne Universit\'eŽs, Laboratoire de Physique Th\'eorique et Hautes
Energies, CNRS UMR 7589,
Universit\'e Pierre et Marie Curie,
4 place Jussieu, 75252 Paris Cedex 05, France}

\date{August 4, 2016}

\begin{abstract}
\noindent We analyze four- and six-derivative
couplings in the low energy effective action of $D=3$ string vacua with half-maximal 
supersymmetry. In analogy with an earlier proposal for the $(\nabla\Phi)^4$ coupling,
we propose that the $\nabla^2(\nabla\Phi)^4$ coupling is  given exactly by a manifestly U-duality
invariant genus-two modular integral. In the limit where a circle
in the internal torus decompactifies, the $\nabla^2(\nabla\Phi)^4$  coupling reduces to 
the $D^2 F^4$ and $\cR^2 F^2$ couplings in  $D=4$, 
along with an infinite series of corrections of order $e^{-R}$, 
from four-dimensional 1/4-BPS dyons whose wordline winds around the circle. Each
of these contributions is weighted by a Fourier coefficient of a meromorphic Siegel modular
form, explaining and extending standard results for the BPS index of 1/4-BPS dyons.
\end{abstract}

\maketitle


String vacua with half-maximal supersymmetry offer an interesting window 
into the non-perturbative regime of string theory and the quantum physics
of black holes, unobstructed by intricacies present in vacua with less
supersymmetry. In particular, the low-energy effective action at two-derivative
order does not receive any quantum corrections, and all higher-derivative
interactions are expected to be invariant under the
action of an arithmetic group $G(\IZ)$, known as the U-duality group,
on the moduli space $G/K$ of massless scalars  \cite{Font:1990gx,Sen:1994fa}.
This infinite discrete symmetry also constrains the spectrum of BPS states, and allows to determine,
for any values of the electromagnetic charges, the number of BPS black hole 
micro-states (counted with signs) in terms of Fourier coefficients of certain 
modular forms \cite{Dabholkar:1989jt,Dijkgraaf:1996xw,Dijkgraaf:1996it}. This property has been used
to confirm the validity of the microscopic stringy description of BPS black holes at an exquisite level
of precision, both for small black holes (preserving half of the supersymmetries
of the background) \cite{Dabholkar:2004yr,Dabholkar:2005dt} and for large black holes (preserving a quarter of the same) \cite{Strominger:1996sh,David:2006yn,Sen:2007qy,Banerjee:2008ky,Banerjee:2011jp,Dabholkar:2012nd,Dabholkar:2014ema,Murthy:2015zzy}. 

In this letter, we shall exploit U-duality invariance and supersymmetry Ward identities to determine 
certain higher-derivative  couplings  in the low-energy effective action of three-dimensional string vacua with 16 supercharges, for all values of the moduli. These protected couplings are analogues of the 
$\cR^4$ and $D^6 \cR^4$ couplings in toroidal compactifications of type II strings,  which have been determined exactly in \cite{Green:1997tv,Green:2005ba} and in many subsequent works.  
Our motivation for studying 
these protected couplings in $D=3$ is that they are expected to encode the infinite spectrum
of BPS black holes in $D=4$, in a way consistent with the U-duality 
group $G_3(\IZ)$. The latter
contains the four-dimensional U-duality group $G_4(\IZ)$, but is potentially far more
constraining. Thus, these protected couplings provide   analogues of `black hole partition functions',  which do not suffer from the usual difficulties in defining thermodynamical partition functions in theories of quantum gravity, and are manifestly automorphic \cite{Gunaydin:2005mx}.

The fact that solitons in $D=4$ 
may induce instanton corrections to the quantum effective potential in dimension $D=3$ 
is well known  in the context of gauge theories with compact $U(1)$ \cite{Polyakov:1976fu}. In the context of quantum field theories with 8 rigid supersymmetries, BPS dyons in four dimensions similarly correct the moduli space metric after
reduction on a circle \cite{Seiberg:1996nz,Gaiotto:2008cd}. 
In  string vacua with 
16 local supersymmetries, one similarly expects that 1/2-BPS dyons in $D=4$ will contribute
to four-derivative scalar couplings of the form 
$F_{abcd}(\Phi)\, \nabla \Phi^a \nabla \Phi^b \nabla \Phi^c
\nabla \Phi^d$ in $D=3$, while both 1/2-BPS
and 1/4-BPS dyons in $D=4$ will contribute to six-derivative scalar couplings of the 
form $G_{ab,cd}(\Phi)\, \nabla (\nabla \Phi^a \nabla \Phi^b ) \nabla (\nabla \Phi^c\nabla \Phi^d)$. 
In either case, the contribution of a four-dimensional BPS state with electric and magnetic charges $(Q,P)$ is expected to be suppressed by $e^{-2\pi R \cM(Q,P)}$, where $\cM(Q,P)$ is the BPS mass and $R$ the radius of the circle on which the four-dimensional theory
is compactified, and weighted by the BPS index $\Omega$ counting the 
number of states with given charges.  
In addition, coupling to gravity implies additional $\cO(e^{-R^2/\ell_P^2})$ corrections from 
gravitational instantons, which are essential for invariance under $G_3(\IZ)$.


For simplicity we shall restrict attention to the simplest three-dimensional string vacuum
with 16 supercharges, obtained by compactifying the ten-dimensional heterotic string on $T^7$.
Our construction can however be generalized with little effort to other models with reduced
rank \cite{Chaudhuri:1995fk}. The moduli space in three dimensions is the symmetric space  
$\cM_3=G_{24,8}$ \cite{Marcus:1983hb}, where $G_{p,q}=O(p,q)/O(p)\times O(q)$ denotes the orthogonal Grassmannian of $q$-dimensional
timelike planes in $\IR^{p,q}$. In the limit where the heterotic string coupling $g_3$ becomes small, 
$\cM_3$ decomposes as 
\be
\label{G3T}
G_{24,8} \to \IR^+
\times G_{23,7} \times \IR^{30}\ ,
\ee
where the first factor corresponds to $g_3$, the second factor to the Narain moduli space (parametrizing the metric, B-field and gauge field on $T^7$),
and $\IR^{30}$ to the scalars $a^I$ dual to the gauge fields in three dimensions.
At each order in $g_3^2$, the low-energy effective action is known to be invariant under 
the T-duality group $O(23,7,\IZ)$, namely the automorphism
group of the Narain lattice $\Lambda_{23,7}$. 
The latter leaves $g_3$ invariant, acts on $G_{23,7}$ by left multiplication and on the last factor
in \eqref{G3T} by the defining  representation. U-duality
postulates that this symmetry is extended to $G_3(\IZ)=O(24,8,\IZ)$, the automorphism group
of the `non-perturbative Narain lattice' $\Lambda_{24,8}=\Lambda_{23,7}\oplus \Lambda_{1,1}$,
where $\Lambda_{1,1}$ is the standard even-self dual lattice of signature $(1,1)$ \cite{Sen:1995wr}. 

In the limit where the radius $R$ of one circle of the internal torus becomes large,  $\cM_3$ instead decomposes as 
\be
\label{G3S}
G_{24,8} \to   \IR^+   \times
\left[ G_{2,1} \times G_{22,6} \right]  \times \IR^{56} \times \IR\ ,
\ee
where the first factor now corresponds to  $R^2/(g_4^2 \ell_H^2)=R/(g_3^2 \ell_H)= R^2/\ell_P^2$ (with $\ell_H$ being
the heterotic string scale, $\ell_P$ the Planck length and and $g_4$ the string coupling in $D=4$), the second correspond to the moduli space $\cM_4$ in 4 dimensions,  the third factor to the holonomies $a^{1I}, a^{2I}$ of the $28$ electric gauge fields fields and their magnetic duals along the circle, and the last factor to the  NUT potential $\psi$, dual to the Kaluza--Klein gauge field. The factor $G_{2,1}\cong SL(2)/U(1)$ is parametrized by the axio-dilaton
$S=S_1+\I S_2=B+\I/g_4^2$, while $G_{22,6}$ is the Narain moduli space of $T^6$, with coordinates $\phi$. 
In the limit $R\to\infty$, the U-duality group is broken to $SL(2,\IZ) \times O(22,6,\IZ)$,
where the first factor $SL(2,\IZ)$ is the famous S-duality in four dimensions \cite{Sen:1994fa,Sen:2005pu,Persson:2015jka}.

Besides being automorphic under $G_3(\IZ)$, the couplings $F_{abcd}$ and $G_{ab,cd}$
must satisfy supersymmetric Ward identities. To state them, we introduce the covariant derivative
$\cD_{ a\hat b}$ on the Grassmannian $G_{p,q}$, defined by its action on the projectors $p^I_{L,a}$
and $p^I_{R,\hat a}$ on the time-like $p$-plane and its orthogonal complement (here and below,
$a,b,$..., $\hat a, \hat b$..., $I,J$... take values $1$ to $p$, $q$, and $p+q$, respectively):
\be
\cD_{a\hat  b} \,p^I_{L,c}= \tfrac12 \delta_{ac} \, p^I_{R,\hat b} \ ,\quad 
\cD_{ a \hat b} \,p^I_{R,\hat c} = \tfrac12 \delta_{\hat b \hat c} \, p^I_{L,a}\ .
\ee
Twice the trace of the operator $\cD^2_{ef} =\cD_{(e}{}^{\hat g} \cD_{f)\hat g}$ is the Laplacian
on $G_{p,q}$. On-shell linearized superspace methods indicate that $F_{abcd}$ and $G_{ab,cd}$ have to satisfy \cite{toappear}
\be
\label{wardf3}
\cD^2_{ef} \, F_{abcd}  =
c_1\,\delta_{ef}\,F_{abcd}
+c_2 \,\delta_{e)(a}\,F_{bcd)(f} 
+c_3\,\delta_{(ab}\,F_{cd)ef} \ ,
\ee
\be
\label{wardg3}
\begin{split}
\cD^2_{ef} G_{ab,cd}=&c_4 \delta_{ef}G_{ab,cd} 
+c_5 \hspace{-0.5mm}  \left[\delta_{e)(a}  G_{b)(f,cd}
+\delta_{e)(c}G_{d)(f,ab}\right]\\
+&c_6 \left[ \delta_{ab}\, G_{ef,cd}+\delta_{cd} \, G_{ef,ab}
-2\delta_{a)(c} \,G_{ef,d)(b} \right] \\
+&c_7 \left[F_{abk(e}\,F^{\phantom{xxx}k}_{f)cd}-F_{c)ka(e}\,F^{\phantom{xxxx}k}_{f)b(d}\right],
\end{split}
\ee
\vspace*{-5mm}
\be
\label{wardfg}
\cD_{[e}{}^{\,[\hat e}  \cD_{f]}{}^{\,\hat f]}  F_{abcd}=0\ ,\quad 
\cD_{[e}{}^{\,[\hat e} \cD_{f}{}^{\,\hat f} \cD_{g]}{}^{\,\hat g]} G_{ab,cd}=0\ .
\ee
The first two constraints are analogous to those derived in  \cite{Lin:2015dsa} for
$H^4$ and $D^2 H^4$ couplings in Type IIB string theory on K3. 
The numerical coefficients $c_1$, ... $\!\! c_7$ will be fixed below from the knowledge of perturbative contributions. 

\section{Exact $(\nabla\Phi)^4$ couplings in $D=3$}

Based on the known one-loop contribution  \cite{Lerche:1987qk,Kiritsis:2000zi}, it was proposed 
in \cite{Obers:2000ta} (and revisited in \cite{Kachru:2016ttg}) 
that the four-derivative coupling $F_{abcd}$
is given exactly by the genus-one modular integral
\be
\label{f4exact}
F^{\scriptscriptstyle (24,8)}_{abcd} = 
\int_{\cF_1}\! \!\!\frac {\de\rho_1\de\rho_2}{\rho_2^{\, 2}} \frac{\partial^4}{
\scalebox{0.9}{$(2\pi\I)^4 \partial y^a
\partial y^b \partial y^c \partial y^d$}}\bigg|_{y=0} \frac{\varGamma_{24,8}}{\Delta}
\ee
where $\cF_1$ is the standard fundamental domain for the action of $SL(2,\IZ)$ on 
the Poincar\'e upper half-plane,
$\Delta=\eta^{24}$ is the unique  cusp form of weight $12$,
and $\varGamma_{24,8}$ is the partition function of the non-perturbative Narain lattice,
\be
\label{defZpq1}
\varGamma_{24,8}=\rho_2^{\, 4}\!\!\!\! \sum_{Q\in\Lambda_{24,8}} \!\!\!\!
e^{\I\pi Q_L^2 \rho - \I \pi Q_R^2 \bar\rho+
2\pi\I Q_L \cdot y+\frac{\pi(y\cdot y)}{2\rho_2} }
\ee
where $Q_L\equiv p_L^I Q_I, Q_R \equiv p_R^I Q_I$, and $|Q|^2=Q_L^2-Q_R^2$ takes
even values on $\Lambda_{24,8}$.
It will be important that the Fourier coefficients of $1/\Delta=\sum_{m\geq -1} c(m)\, q^m$
count the number of 1/2-BPS states in the $D=4$ vacuum obtained by decompactifying a circle inside $T^7$. This is obvious from the fact that these states are dual to perturbative string states  carrying only left-moving excitations \cite{Dabholkar:1989jt,Dabholkar:2004yr}.
It is also worth noting that the ansatz \eqref{f4exact} is a special case of a more general class of one-loop integrals,  which we shall denote by $F_{abcd}^{\scriptscriptstyle (q+16,q)}$, where the lattice $\Lambda_{24,8}$
is replaced by an even self-dual lattice $\Lambda_{q+16,q}$ and the factor $\rho_2^4$ by $\rho_2^{q/2}$. The integral $F_{abcd}^{\scriptscriptstyle (q+16,q)}$ converges for $q<6$, and is defined for $q\geq 6$ by a suitable renormalization prescription. For any value of $q$, $F_{abcd}^{\scriptscriptstyle (q+16,q)}$ satisfies \eqref{wardf3} and
\eqref{wardfg} with $c_1=\tfrac{2-q}{4}, \ c_2=4-q, \ c_3=3$. 

By construction, the ansatz \eqref{f4exact} is a solution of the  supersymmetric Ward identity,
which is manifestly invariant under $G_3(\IZ)$. Its expansion at weak coupling (corresponding to 
the parabolic decomposition \eqref{G3T}, such that the non-perturbative Narain lattice 
$\Lambda_{24,8}$ degenerates to $\Lambda_{23,7}\oplus \Lambda_{1,1}$) can be computed 
using the standard unfolding trick. For simplicity, we shall assume that none of the indices $abcd$ lies along $\Lambda_{1,1}$:
\be
\label{f3expT}
\begin{split}
F^{\scriptscriptstyle (24,8)}_{\alpha\beta\gamma\delta} =&  \frac{c_0}{16\pi g_3^{\, 4}} \, \delta_{(ab}\delta_{cd)} 
+ \frac{F^{\scriptscriptstyle (23,7)}_{\alpha\beta\gamma\delta}}{g_3^{\, 2}} 
+ 4 \sum_{k=1}^3 \sum_{Q\in\Lambda^\star_{23,7}} \!\! P^{(k)}_{\alpha\beta\gamma\delta} \\
 \times  \bar c(Q) &\, g_3^{2k-9}\, |\sqrt2 Q_R|^{k-\tfrac72}\, 
K_{k-\tfrac72}\left(\tfrac{2\pi}{g_3^{\, 2}} |\sqrt{2}Q_R|\right)\,
e^{-2\pi\I a^I Q_I}
\end{split}
\ee  
where $c_0=24$ is the constant term in $1/\Delta$, $\Lambda^\star=\Lambda\backslash\{0\}$, 
$P_{\alpha\beta\gamma\delta}^{(1)}(Q)=Q_{L\alpha} Q_{L\beta} Q_{L\gamma} Q_{L\delta}$, 
$P_{\alpha\beta\gamma\delta}^{(2)}= -\tfrac{3}{2\pi} \delta_{(\alpha\beta} Q_{L\gamma} Q_{L\delta)}$, 
$P_{\alpha\beta\gamma\delta}^{(3)} = \tfrac{3}{16\pi^2} \delta_{(\alpha\beta} \delta_{\gamma\delta)}$,
$K_\nu(z)$ is the modified Bessel function of the second kind, behaving as $\sqrt{\frac{\pi}{2z}} e^{-z}
(1+\cO(1/z))$
for large positive values of $z$,
and 
\be
\bar c(Q) = \sum_{d|Q} d\, c\bigl( -\tfrac{|Q|^2}{2d}\bigr) \ .
\ee
After rescaling from Einstein frame to string frame, the first and second terms in \eqref{f3expT} are recognized as the tree-level and one-loop $(\nabla \Phi)^4$ coupling in perturbative heterotic
string theory, while the remaining terms correspond to NS5-brane and KK5-branes wrapped on any possible $T^6$ inside $T^7$ \cite{Obers:2000ta}. 

In the large radius limit (corresponding to 
the parabolic decomposition \eqref{G3S}, such that the non-perturbative Narain lattice 
$\Lambda_{24,8}$ degenerates to $\Lambda_{22,6}\oplus \Lambda_{2,2}$), we get instead
(in units where $\ell_P=1$)
\bea
\label{f3expS}
F^{\scriptscriptstyle (24,8)}_{\alpha\beta\gamma\delta} \hspace{-0.8mm}&=&\hspace{-0.8mm} R^2 \left( \frac{c_0}{16\pi} \, \widehat  E_1(S) \, \delta_{(\alpha\beta}\delta_{\gamma\delta)} 
+  F^{\scriptscriptstyle (22,6)}_{\alpha\beta\gamma\delta} \right) 
\\
&&\hspace{-4.8mm} + 4 R^2\sum_{k=1}^3 \sum_{Q'\in\Lambda^\star_{22,6}}  \sum'_{m,n}\, 
c\bigl( -\tfrac{|Q'|^2}{2}\bigr)  \, P^{(k)}_{\alpha\beta\gamma\delta} \nonumber \\
&& \hspace{-5.0mm} K_{k-\tfrac72}\left(\tfrac{2\pi R |mS+n|}{\sqrt{S_2}} |\sqrt{2}Q'_R|\right)\,
e^{-2\pi\I (m a^1+ n a^2)\cdot Q'}\hspace{-1.2mm} + \dots \nonumber 
\eea  
where $\widehat E_1(S)=-\tfrac{3}{\pi}\,\log S_2 |\eta(S)|^4$.
The first term in \eqref{f3expS} originates from the dimensional reduction of the $\cR^2$ and 
$F^4$ couplings in $D=4$ \cite{Harvey:1996ir,Kiritsis:2000zi}, 
after dualizing the gauge fields into scalars.  
The second term  in \eqref{f3expS} is of order $e^{-2\pi R \cM(Q,P)}$, 
where $\cM$ is the mass of a four-dimensional 1/2-BPS state with electromagnetic charges
$(Q,P)=(mQ',nQ')$. The phase factor is the expected minimal coupling of the dyonic state to the holonomies of the electric and magnetic gauge fields along the circle. Fixing 
charges $(Q,P)$ such that $Q$ and $P$ are collinear, the sum over $(m,n)$ induces a measure 
factor 
\be
\mu(Q,P) = \sum_{d|(Q,P)} c\bigl( -\tfrac{{\rm gcd}(Q^2, P^2, Q\cdot P)}{2d^2}\bigr)\ ,
\ee
which is  recognized as the degeneracy of 1/2-BPS states with charges $(Q,P)$. In particular for a purely electric state ($P=0$) with primitive charge, it reduces to $c(-|Q|^2/2)$.
The dots in   \eqref{f3expS} stand for terms of order $e^{-2\pi R^2|k|+2\pi\I k\psi}$,
characteristic of a Kaluza--Klein monopole of the form ${\rm TN}_k \times T^6$, where 
${\rm TN}_k$ is Euclidean Taub--NUT space with  charge $k$. These contributions will be 
discussed in \cite{toappear}. 

\section{Exact $\nabla^2(\nabla\Phi)^4$ couplings in $D=3$}

We now turn to the six-derivative coupling $G_{ab,cd}$, which is expected to receive both
1/2-BPS and 1/4-BPS instanton contributions. Based on U-duality invariance, supersymmetric
Ward identities and the known two-loop contribution \cite{D'Hoker:2005jc,D'Hoker:2005ht}, 
it is natural to conjecture that $G_{ab,cd}$ is given by the genus-two modular integral  
\be
\label{d2f4exact0}
G^{\scriptscriptstyle (24,8)}_{ab,cd}= 
\int_{\cF_2} \!\!\!\frac{\de^3\Omega_1\de^3\Omega_2}{|\Omega_2|^3} 
\frac{\tfrac{1}{2}(  \scalebox{0.9}{$ \varepsilon_{ik} \varepsilon_{jl}+ \varepsilon_{il} \varepsilon_{jk} $})\partial^4 }
{\scalebox{0.9}{$(2\pi\I)^4\partial y^a_i \partial y^b_j
 \partial y^c_k \partial y^d_l $}}\bigg|_{y=0} \frac{\varGamma_{24,8,2}}{\Phi_{10}}\ ,
\ee
where $\cF_2$ is the standard fundamental domain for the action of $Sp(4,\IZ)$ on the Siegel upper half-plane of degree two \cite{zbMATH03144647}, $|\Omega_2|$ is the determinant of the imaginary
part of $\Omega=\Omega_1+\I\Omega_2$,
$\Phi_{10}$ is the unique cusp form of weight 10 under the Siegel modular group 
$Sp(4,\IZ)$ (whose inverse counts micro-states of 1/4-BPS black holes \cite{Dijkgraaf:1996it}), 
and $\varGamma_{24,8,2}$ is the genus-two partition function of the
non-perturbative Narain lattice, 
\be
\label{defZpq2}
\varGamma_{24,8,2} \hspace{-0.4mm} = \hspace{-0.6mm} |\Omega_2|^{4}\hspace{-3mm}\sum_{Q^i \in\Lambda_{24,8}^{\otimes 2}} \hspace{-3mm}
e^{\I\pi (Q_L^i \Omega_{ij} Q_L^j  -  Q_R^i \bar\Omega_{ij} Q_R^j +2 Q_L^i y_i) +\pi y_i^a \Omega_2^{ij} y_{ja}}
\ee
We shall denote by $G_{ab,cd}^{\scriptscriptstyle (q+16,q)}$ the analogue of \eqref{defZpq2} where
the lattice $\Lambda_{24,8}$ is replaced by $\Lambda_{q+16,q}$ and the power of $|\Omega_2|$
by $q/2$. The integral $G_{ab,cd}^{\scriptscriptstyle (q+16,q)}$ is convergent for $q<6$, and defined for $q\geq 6$ by a suitable renormalization prescription \cite{Pioline:2015nfa}.
For any value of  $q$, one can show that 
$G_{ab,cd}^{\scriptscriptstyle (q+16,q)}$ satisfies \eqref{wardg3} and \eqref{wardfg} with $c_4=\tfrac{3-q}{2}, 
c_5=\tfrac{6-q}{2}, 
\ c_6=\tfrac12,\ c_7=-\pi$. In particular, the quadratic source term on the r.h.s. of 
\eqref{wardg3} follows from the pole of $1/\Phi_{10}$ on the separating degeneration divisor,
similar to the analysis in \cite{Pioline:2015nfa}. Thus, $G^{(24,8)}$ is a solution of the  supersymmetric Ward identity, which is manifestly invariant under $G_3(\IZ)$. It remains to check that it produces the expected terms at weak coupling, when $\Lambda_{24,8}$ degenerates to $\Lambda_{23,7}\oplus \Lambda_{1,1}$. This limit can be studied using a higher-genus version of the unfolding trick \cite{Pioline:2014bra,Florakis:2016boz}. Using results about the Fourier--Jacobi expansion of $1/\Phi_{10}$ from \cite{Dabholkar:2012nd},  we find
\bea
\label{g3expT}
G^{\scriptscriptstyle (24,8)}_{\alpha\beta,\gamma\delta} \hspace{-1.5mm} &=& \hspace{-1.5mm}  \frac{G^{\scriptscriptstyle (23,7)}_{\alpha\beta,\gamma\delta}}{g_3^{\, 4}}  -\frac{\delta_{\alpha\beta}  G^{\scriptscriptstyle (23,7)}_{\gamma\delta}\hspace{-0.7mm}+\hspace{-0.5mm}\delta_{\gamma\delta}  G^{\scriptscriptstyle (23,7)}_{\alpha\beta}\hspace{-0.7mm}-\hspace{-0.5mm} 2 \delta_{\gamma)(\alpha}  G^{\scriptscriptstyle (23,7)}_{\beta)(\delta} }{12g_3^{\, 6}} 
\nonumber \\
&& - \frac{1}{2\pi g_3^{\, 8}} \left[\delta_{\alpha\beta} \delta_{\gamma\delta} 
- \delta_{\alpha(\gamma} \delta_{\delta)\beta} \right]+ \dots
\eea
where
\be
G_{ab}^{\scriptscriptstyle (q+16,q)} = \int_{\cF_1}\! \!\!\frac{\de\rho_1\de\rho_2}{\rho_2^{\, 2}}
\frac{\partial^2}{\scalebox{0.8}{$(2\pi\I)^2 \partial y^a
\partial y^b$}}\Big|_{y=0} \frac{\widehat E_2\, \varGamma_{q+16,q}}{\Delta}\ ,
\ee
with $\widehat E_2=\tfrac{12}{\I\pi}\partial_\rho\log\eta-\tfrac{3}{\pi\rho_2}$ the almost holomorphic Eisenstein series of weight 2.
The first and second terms in \eqref{g3expT} corresponds to the zero and rank 1 orbits, respectively. 
The third term is missed by a naive unfolding procedure, which fails due to the singularity of the integrand in the separating degeneration limit, but is crucial to ensure consistency with the supersymmetric Ward identity \eqref{wardg3}.  After rescaling to string frame, the first three terms in \eqref{g3expT} correspond
to the expected two-loop \cite{D'Hoker:2005jc,D'Hoker:2005ht}, one-loop  \cite{Sakai:1986bi}  and tree-level contributions \cite{Gross:1986mw,Bergshoeff:1989de} to the $\nabla^2(\nabla\Phi)^4$ coupling in heterotic string on $T^7$, while  the dots stand for terms of order $e^{-1/g_3^2}$ ascribable to NS5-brane and KK5-brane instantons, which will be discussed in \cite{toappear}. Note that  the tree-level single trace $D^2 F^4$ term in \cite{Gross:1986mw} proportional to $\zeta(3)$ vanishes on the Cartan subalgebra \cite{Drummond:2003ex}, and does not contribute to this coupling. 

Having shown that our ansatz \eqref{d2f4exact0} passes all consistency conditions in $D=3$,
let us now analyze its large radius limit, where $\Lambda_{24,8}$ degenerates to 
$\Lambda_{22,6}\oplus \Lambda_{2,2}$. Again, the unfolding trick gives
\bea
\label{g3expS}
&& G^{\scriptscriptstyle (24,8)}_{\alpha\beta,\gamma\delta} \\
 &=& \hspace{-1mm}R^4 \Bigl[ \scalebox{0.9}{$ G^{\scriptscriptstyle (22,6)}_{\alpha\beta,\gamma\delta}  - 
 \scalebox{1.2}{$\frac{\widehat{E}_1(S)}{12} $}
 \left( \delta_{\alpha\beta}  G^{\scriptscriptstyle (22,6)}_{\gamma\delta}\hspace{-0.7mm}+\hspace{-0.5mm}\delta_{\gamma\delta}  G^{\scriptscriptstyle (22,6)}_{\alpha\beta}\hspace{-0.7mm}-\hspace{-0.5mm} 2 \delta_{\gamma)(\alpha}  G^{\scriptscriptstyle (22,6)}_{\beta)(\delta}  \right) $} \Bigr.\nonumber \\
&&+ g(S)  \Bigl. (\scalebox{0.9}{$ \delta_{\alpha\beta} \delta_{\gamma\delta}-\delta_{\alpha(\gamma} \delta_{\delta)\beta} $}) 
\Bigr] 
\scalebox{0.9}{$+\, G^{(1)}_{\alpha\beta,\gamma\delta} + G^{(2)}_{\alpha\beta,\gamma\delta} 
+ G^{(\rm KKM)}_{\alpha\beta,\gamma\delta}$} \nonumber
\eea
The two terms on the first line (which correspond to  the constant term  with respect to the parabolic decomposition \eqref{G3S}) originate from the reduction of the $D^2 F^4$ and $\cR^2 F^2$ couplings
in four dimensions. The term proportional to $g(S)$ originates from the separating degeneration
divisor, and will be determined in \cite{toappear}. The terms $G^{(1)}$ and $G^{(2)}$ are independent of the NUT potential $\psi$, and correspond to the Abelian Fourier coefficients. They are both suppressed as $e^{-2\pi R\cM(Q,P)}$, but $G^{(1)}$ has support on electromagnetic charges $(Q,P)$ which $Q$ and $P$ collinear, hence corresponds to contributions of 1/2-BPS states  winding the circle, while $G^{(2)}$ has support on generic charges, corresponding to 1/4-BPS states. The last term $G^{(\rm KKM)}$ includes all terms with non-zero charge with respect to the NUT potential, corresponding to Kaluza--Klein monopole contributions. 

In this letter, we focus on the contribution $G^{(2)}$ from 1/4-BPS black holes. This contribution originates from the `Abelian rank 2 orbit', whose stabilizer is the parabolic subgroup $GL(2,\IZ)\ltimes \IZ^3$ inside $Sp(4,\IZ)$. Thus, the integral can be unfolded onto $\cP_2/GL(2,\IZ) \times [0,1]^3$,
where $\cP_2$ denotes the space of positive definite $2\times 2$ matrices $\Omega_2$:
\be
\begin{split}
G^{(2)}_{\alpha\beta,\gamma\delta}  =& R^4\, \int_{\cP_2} \frac{\de^3\Omega_2}{|\Omega_2|^3}
 \, \int_{[0,1]^3} \!\!\!\!\!\! \de^3\Omega_1\, 
\frac{( \scalebox{0.9}{$\varepsilon_{ik} \varepsilon_{jl}+ \varepsilon_{il} \varepsilon_{jk} $})\partial^4}{\scalebox{0.9}{$(2\pi\I)^4\partial y^a_i \partial y^b_j \partial y^c_k \partial y^d_l $}}\bigg|_{y=0}
\\
 & \hspace*{-12.6mm}
 \times\frac{\bigl\langle e^{\scalebox{0.7}{$-2\pi\I a^{iI} A_{ij} Q^j_I$}} \bigr\rangle_{\scriptscriptstyle 22,6,2}}{ \Phi_{10}} \hspace{-6.5mm} \sum_{A\in M_2(\IZ) /GL(2,\IZ)
\atop  |A|\ne 0}\hspace{-6mm}
e^{-\frac{\pi R^2}{S_2} \Tr\Bigl[\Omega_2^{-1} \cdot A^\intercal \cdot {\scriptsize \begin{pmatrix} 1 & S_1 \\ S_1 & |S|^2 \end{pmatrix} }\cdot A\Bigr]}
\end{split}
\ee
\vspace{-3mm}\\
where $\langle f(Q) \rangle_{22,6,2}$ denotes the partition function $\Gamma_{22,6,2}$
with an insertion of $f(Q)$ in the sum.
The integral over $\Omega_1$ at fixed $\Omega_2$ extracts the Fourier coefficient $
C\left[{\scriptsize \left(\begin{array}{cc} - \tfrac12|Q|^2 & -Q\cdot P \\ -Q\cdot P & -\tfrac12|P|^2 \end{array}\right)};\Omega_2\right]$ of $1/\Phi_{10}$. Due to the zeros of $\Phi_{10}$, the latter 
is a locally constant function of $\Omega_2$, discontinuous across certain real codimension 1 walls 
in $\cP_2$ \cite{Sen:2007vb,Dabholkar:2007vk}. For large $R$ however, the remaining integral over $\Omega_2$ is dominated by a saddle point $\Omega_2^\star$ (see \eqref{Omsaddle} below), so to all orders in $1/R$ around the saddle point, we can replace the above Fourier coefficient by its value at  $\Omega_2^\star$. The remaining integral over $\Omega_2$ can be computed
using 
\be
\label{SintB}
\int_{\cP_2}\de^3 S\,|S|^{\delta-\frac{3}{2}}\, e^{ -\pi{\rm Tr}\,(SA+S^{-1}B)}\,
=2\left(\frac{|B|}{|A|}\right)^{\delta/2}\,\widetilde B_\delta(AB)\ ,
\ee
where $\widetilde B_\delta(Z)$ is a matrix-variate generalization of the modified Bessel function \cite{0066.32002}\footnote{Our $\widetilde B_\delta(Z)$ is related to $B_\delta(Z)$ in \cite{0066.32002} via $\widetilde B_\delta(Z/\pi^2)=\frac12 (\det Z)^{\delta/2} B_\delta(Z)$.}, which depends on $Z$ only through its trace and determinant. In the limit where both are large, one has
\be
\label{BKlim}
\widetilde{B}_\delta(Z) \sim  |Z|^{-1/4}\, K_0\left( 2\pi \sqrt{\Tr Z+2\sqrt{|Z|}}\right)\ .
\ee
Further relabelling $({Q\atop P})=
A  ({Q_1 \atop Q_2})$, we find 
\bea
 \label{F2ranktwoAb}
G^{(2)}_{\alpha\beta,\gamma\delta}\hspace{-1.1mm}Ê  &=& \hspace{-0.9mm}Ê2R^{7} \hspace{-1.8mm}\sum_{Q,P\in\Lambda^\star_{22,6}}\hspace{-2.5mm}e^{-2\pi \I (a^1 Q+a^2 P)} 
\mu(Q,P) 
 \nonumber \\
&& \hspace{-15mm} \times \sum_{k=1}^3 \frac{P^{(k)}_{\alpha\beta,\gamma\delta}}{\scalebox{0.8}{$|P_R\wedge Q_R|^{\frac{4-k}{2}}$} }   \widetilde  B_{\frac{4-k}{2}} \!  \left[\tfrac{2R^2}{S_2}
{\scalebox{0.7}{$ \begin{pmatrix} 1 & \! S_1 \\ S_1 &\!  |S|^2 \end{pmatrix}$} } {\scalebox{0.7}{$ \begin{pmatrix} |Q_R|^2 & \! P_R \cdot Q_R \\ P_R \cdot Q_R & |P_R|^2
\end{pmatrix}$}}
 \right] \qquad
\eea
where $|P_R\wedge Q_R|=\sqrt{(P_R^2)(Q_R^2)-(P_R\cdot Q_R)^2}$, $P^{(k)}_{\alpha\beta,\gamma\delta}$ is a polynomial of degree $6-2k$ in $Q_L$,
\be
\label{mu14}
\mu(Q,P)=  \hspace{-5.5mm} \sum_{\substack{{A}\in M_2(\IZ)/GL(2,\IZ) \\ 
 A^{-1}({Q\atop P})\in \Lambda_{22,6}^{\otimes 2}}}  \hspace{-5.9mm} 
|A|\, 
C\left[A^{-1} {\scriptsize \left(\begin{array}{cc} - \tfrac12|Q|^2 & \! -Q\cdot P \\ -Q\cdot P &\!  -\tfrac12|P|^2 \end{array}
\right)} A^{-\intercal};\Omega_2^\star\right]
\ee
and $\Omega_2^\star$ is the location of the afore-mentioned saddle point,
\be
\label{Omsaddle}
\Omega_2^\star\hspace{-0.5mm} =\hspace{-0.5mm} \frac{R}{\scalebox{0.8}{$ \cM(Q,P)$}} A^\intercal \hspace{-1mm}  \left[ \tfrac{1}{S_2} {\scalebox{0.7}{$  \begin{pmatrix} 1 & S_1 \\ S_1 & |S|^2 \end{pmatrix} $}}
\hspace{-0.2mm}  \scalebox{0.9}{$+$} \hspace{-0.2mm} \scalebox{0.9}{$\tfrac{1}{|P_R \wedge Q_R|} $} 
{\scalebox{0.7}{$ \begin{pmatrix} |P_R|^2 & \hspace{-1mm}-P_R \cdot Q_R \\ -P_R \cdot Q_R & |Q_R|^2
 \end{pmatrix}$ }}
\right] \hspace{-1mm}  A\, .
\ee
Using \eqref{BKlim}, we see that these contributions behave as 
$e^{-2\pi R \cM(Q,P)}$ in the limit $R\to \infty$, where 
\be
\cM(Q,P)=\sqrt{ 2 \tfrac{|Q_R-S P_R|^2}{S_2} + 4 \sqrt{ \left|  
{\scriptsize \begin{matrix} |Q_R|^2 & Q_R \cdot P_R \\ Q_R \cdot P_R & |P_R|^2
 \end{matrix} } \right| }}
\ee
is recognized as the  mass of a 1/4-BPS dyon
with electromagnetic charges $(Q,P)$ \cite{Cvetic:1995uj,Duff:1995sm}. Moreover, 
in cases where only $A=\mathds{1}$ contributes 
to \eqref{mu14}, the instanton measure $\mu(Q,P)$ agrees 
with the BPS index $\Omega(Q,P;S,\phi)$ in the corresponding chamber of the moduli space 
$\cM_4$ in $D=4$, computed 
with the contour  prescription in \cite{Cheng:2007ch}. Our result \eqref{mu14} generalizes
this prescription to arbitrary electromagnetic charges $(Q,P)$ and recovers
the results of \cite{Banerjee:2008pv,Banerjee:2008pu,Dabholkar:2008zy} for dyons with 
torsion, fixing a subtlety in the choice of chamber. Additional (exponentially suppressed) contributions to
$G^{(2)}$ arise from the difference between $C\left[{\scriptsize \left(\begin{array}{cc} - \tfrac12|Q|^2 & -Q\cdot P \\ -Q\cdot P & -\tfrac12|P|^2 \end{array}\right)};\Omega_2\right]$ and its value at the saddle point. The relation between the jumps of these Fourier coefficients and the possible splittings of a 
1/4-BPS bound state into two 1/2-BPS constituents \cite{Sen:2007vb} is crucial for consistency with the quadratic source term in the supersymmetric Ward identity \eqref{wardg3}. 
These contributions, along with the terms $G^{(2)}$ and $G^{(\rm KKM)}$ which we have ignored here, will be discussed in \cite{toappear}. 

\section{Discussion}

In this work, we have determined the exact $(\nabla\Phi)^4$ and $\nabla^2(\nabla\Phi)^4$
couplings in the low energy effective action of $D=3$ string vacua with half-maximal supersymmetry, 
focussing on the simplest model, heterotic string compactified on $T^7$. Our ans\"atze 
\eqref{f4exact} and \eqref{d2f4exact0} are manifestly U-duality invariant,  satisfy the requisite supersymmetric Ward identities, and reproduce the known perturbative contributions at weak
heterotic coupling. In the limit where the radius of one circle inside $T^7$ becomes large, they 
yield the exact $F^4, \cR^2, D^2 F^4$ and $\cR^2 F^2$ couplings in $D=4$, plus an infinite
series of corrections of order $e^{-2\pi R \cM(Q,P)}$ which are interpreted as Euclidean counterparts of four-dimensional BPS states with mass $\cM(Q,P)$, whose worldline winds around the circle. Quite remarkably, the contribution from  a 1/4-BPS dyon is weighted by the BPS index $\Omega(Q,P;S,\phi)$, extracted from the Siegel modular form $1/\Phi_{10}$ using the contour prescribed in  \cite{Cheng:2007ch}. Indeed, it was suggested in \cite{Gaiotto:2005hc} (see also \cite{Dabholkar:2006bj,Dabholkar:2006xa}) to represent 1/4-BPS dyons as heterotic strings wrapped on a genus-two curve holomorphically embedded in a $T^4$ inside $T^7$. This picture was further used in \cite{Banerjee:2008yu} to justify the contour prescription of \cite{Cheng:2007ch}. Our analysis of the $\nabla^2(\nabla\Phi)^4$ coupling in $D=3$ 
gives a concrete basis to these heuristic ideas, and explains why 
1/4-BPS dyons in $D=4$ are counted by a Siegel modular form of genus two. A more
detailed analysis of the weak coupling and large radius expansions of the $\nabla^2(\nabla\Phi)^4$ coupling will appear in \cite{toappear}, with particular emphasis on the consequences of wall-crossing 
for three-dimensional couplings.

\medskip\medskip

\noindent  {\it Acknowledgments:} We are grateful to Eric d'Hoker, Ioannis Florakis and Rodolfo Russo for valuable discussions on genus-two modular integrals, and to Sameer Murthy for discussions on wall-crossing in $\cN=4$ string vacua. G.B. and C.C.H. thank CERN for its hospitality. 


\end{document}